*Title Page*

# Bridging Photon Statistics and Phase Transitions in Random Fiber Lasers

Yifei Qi[1], Runhao Li[1], Jie Li[1], Taichao Wang[1], Wangyouyou Li[1], Ernesto P. Raposo[2], Anderson S. L. Gomes[3], Han Wu[4] and Zinan Wang[1,*]

[1] *Key Lab of Optical Fiber Sensing & Communications, University of Electronic Science and Technology of China (UESTC), Chengdu, China*

[2] *Laboratório de Física Teórica e Computacional, Departamento de Física, Universidade Federal de Pernambuco, 50670-901 Recife, Pernambuco, Brazil*

[3] *Departamento de Física, Universidade Federal de Pernambuco, Recife, Pernambuco, Brazil*

[4] *College of Electronics and Information Engineering, Sichuan University, Chengdu, 610000, Sichuan, China.*

Yifei Qi, Email: YFQi98@outlook.com

Runhao Li, Email: 202421011617@std.uestc.edu.cn

Jie Li, Email: 202421011615@std.uestc.edu.cn

Taichao Wang, Email: 202421011616@std.uestc.edu.cn

Wangyouyou Li, Email: 202422011618@std.uestc.edu.cn

Ernesto P. Raposo, Email: ernesto.raposo@ufpe.br

Anderson S. L. Gomes, Email: andersonslgomes@gmail.com

Han Wu, Email: hanwu@scu.edu.cn

Zinan Wang **(Corresponding author)**, Email: znwang@uestc.edu.cn

# Abstract

**Complex systems exhibit rich equilibrium states, yet the universal principles governing these systems remain unrevealed, motivating the search for novel experimental platforms. Random fiber lasers (RFLs), which generate partially-coherent light-wave through feedback from Rayleigh scattering, provide a photonic realization of such systems. Here we report a comprehensive theoretical and experimental investigation of photon statistics for RFLs based on classical second-order temporal correlation function $g^{(2)}(\tau)$, revealing unique statistical properties and introduce a two-dimensional framework for controlling photon statistics. Remarkably, we establish a unified landscape between photon correlation, intensity statistics governed by Lévy statistics, and phase transitions with replica symmetry breaking. This multifaceted relationship, observed for the first time, bridges disordered photonics with statistical physics of complex system. Our results offer new pathways for engineering laser emission with controllable photon statistics, and more broadly, this work positions RFLs as a fertile land for exploring emergent behaviors in disordered systems.**

# Introduction

An essential feature of complex systems is the presence of numerous metastable states, which give rise to rich complex dynamics such as replica symmetry breaking and glass-like behavior[1-4]. Uncovering the universal principles that govern these emergent behaviors remains a central challenge in contemporary physics, and identifying experimental platforms that are both controllable and capable of exhibiting complex dynamics is crucial to meeting this challenge[4,5]. In recent years, random fiber lasers (RFLs) based on disordered media have emerged as a promising photonic platform for studying the statistical physics of complex systems, owing to their intrinsic coupling of disorder, nonlinearity, and complex dynamics[6-9]. Unlike conventional lasers that rely on fixed cavities, or amplified spontaneous emission (ASE) sources that operate without cavity feedback, RFLs utilize multiple photon-scattering within disordered media as their feedback mechanism (for reviews on random lasers and RFLs, see Ref.10). Their dynamics is further enriched by stimulated Raman scattering (SRS), four-wave mixing (FWM), and other nonlinear processes whose cooperation and competition collectively construct a rugged energy landscape with multiple local minima, thereby enabling the exploration of complex-system behaviors[11-16]. Beyond their fundamental significance, RFLs also offer practical advantages—including compact architecture, high efficiency, and spectral tunability[17-19], that have enabled applications in sensing, imaging, and optical communication[20-25]. Their intrinsically broad spectral characteristics further position them as promising seed sources for laser-driven inertial confinement fusion (ICF) systems[26]. Moreover, RFLs have been employed to observe and investigate a variety of macroscopic complex phenomena, including photonic phase transitions, turbulence, and photonic Hall effects, establishing its unique position as an experimental platform for looking into the statistical physics of complex system phenomena[27-34].

As a key parameter characterizing the statistical properties of radiation sources, the second-order temporal correlation function $g^{(2)}(\tau)$ has been widely employed across various frontier scientific domains[35-38]. In optics, second-order temporal correlation function serves not only as a fundamental tool for probing the intrinsic statistical nature of light sources, but also plays a crucial role in understanding the complex field dynamics and intensity fluctuation statistics.

Despite its significance, the study of second-order temporal correlation in RFLs remains at preliminary stage. In the context of three-dimensional random lasers, H. Cao et al. investigated the photon statistics of ZnO-based random lasers and revealed that the output coherence can be effectively tuned by varying the pump power[39]. Building on this, a master-equation-based theoretical framework was introduced to further elucidate the interplay between scattering strength and pumping rate, highlighting the impact of spatial inhomogeneity on photon dynamics[40]. More recently, efforts have been directed toward exploring second-order temporal correlation in RFL systems[41,42]. Raposo et al. developed a theoretical method based on random matrix theory to compute the second-order temporal correlation function of RFLs[42]. In parallel, they experimentally evaluated $g^{(2)}(\tau)$ in an erbium-doped RFL by analyzing temporal intensity sequences, confirming both the validity of the theoretical model and its connection to the eigenvalue statistics of random matrices.

Although previous studies have provided valuable theoretical insights, a systematic understanding of the interplay between the photon statistics and phase transitions in RFLs still faces three major challenges. First, the temporal resolution of existing detection systems remains insufficient to match the coherence time of the laser source, hindering accurate observation of rapid intensity fluctuations. Second, the evolution and regulation of macroscopic statistical parameters in RFLs (such as $g^{(2)}(\tau)$, Lévy index $\alpha$, and the replica overlap parameter $q$) lack a unified analytical framework. Third, the fundamental connection between photon statistics and photonic phase transitions has not yet been established, leaving the theoretical description of RFLs as complex systems incomplete.

To address these challenges, this work presents a comprehensive theoretical and experimental investigation for the photon statistics of RFL by the second-order temporal correlation function. By systematically investigating the evolution of $g^{(2)}(\tau)$ in three types of light sources with distinct feedback mechanisms (conventional single-frequency fiber laser, RFL, and ASE), the unique photon statistics behavior intrinsic to RFL is revealed. Building upon this, a tunability mechanism based on pump power and the Rayleigh scattering phase state is proposed, enabling effective tunability of the photon statistics, which opens up promising avenues for developing photon sources with tunable statistical properties for applications such as seed sources in ICF. Notably, this study establishes, for the first time, a fundamental connection between second-order temporal correlations, intensity statistics (represented by Lévy statistics), and phase transition phenomena (represented by the intensity fluctuation replica overlap parameter) in RFLs, across the dual dimensions of pump power and scatterer state. These findings not only deepen the understanding of intrinsic laser dynamics in disordered media, but also provide critical insights for advancing interdisciplinary research in disordered photonics, statistical physics, and complex systems.

## Theoretical analysis

Conventional fiber lasers, ASE sources, and RFLs represent three typical classes of fiber-based light sources, whose fundamental differences mainly originate from the degree of order in their feedback mechanisms. Through a systemic analysis of their macroscopic parameters, this study reveals that these light sources exhibit distinctly different energy landscapes, as illustrated in

Fig. 1.

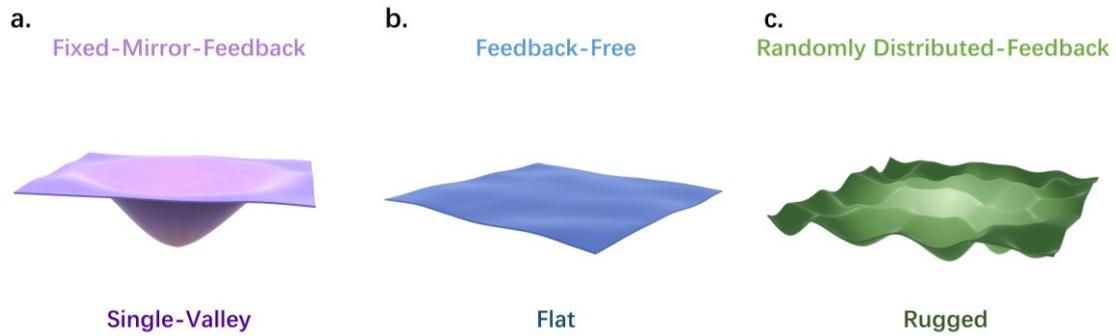

**a.** Fixed-Mirror-Feedback     **b.** Feedback-Free     **c.** Randomly Distributed-Feedback

Single-Valley     Flat     Rugged

**Figure 1. Energy landscape of three typical classes of fiber-based light sources.** a. Fixed-Mirror-Feedback (single-frequency fiber laser): single-valley. b. Feedback-free (ASE source): flat. c. Randomly distributed-feedback (RFL): rugged.

Conventional fiber lasers rely on fixed cavity mirrors to provide deterministic feedback. Their modal degrees of freedom are limited, the system dynamics can be captured by a few variables, and the energy landscape is essentially single-valley, leading to stable oscillatory operation, as shown in Fig. 1a. ASE sources, lacking any resonant feedback, exhibit completely random modal phases without effective coupling. They behave as thermal-like sources with a flat energy landscape and do not support photonic phase transitions, as shown in Fig. 1b.

In contrast, RFLs employ randomly distributed feedback arising from randomly distributed Rayleigh scattering in optical fiber, strongly coupled with nonlinear processes such as stimulated Raman scattering. The cooperative interplay of disorder, nonlinearity, and amplification gives rise to a multivalley and rugged energy landscape whose structure adaptively reconstructs with the pump and scatterer-state dynamics. When the pump power is below threshold, the RFL operates in a spontaneous-emission regime characterized by a nearly flat energy landscape. Once the pump power exceeds the threshold, the RFL develops a multivalley energy landscape, as shown in Fig. 1c. Numerous modes compete and cooperate within this rugged landscape, generating strong frustration and endowing RFLs with the statistical features of complex systems. Moreover, this study reveals that as the fluctuation-degree of Rayleigh-scattering phase increases, the number of local minima in the multivalley energy landscape gradually decreases and ultimately collapses into a flattened landscape.

Within this framework, three macroscopic statistical parameters enable a unified landscape of photon statistics and phase transitions in RFLs:

(i) Second-order temporal coherence $g^{(2)}(\tau)$, describing the dynamical correlations of intensity fluctuations, which can be expressed as:

$$g^{(2)}(\tau) = \frac{\langle I(t+\tau)I(t)\rangle}{\langle I(t+\tau)\rangle\langle I(t)\rangle} \tag{1}$$

where $I(t)$ is the emission intensity of the laser source at the time t and <...> means the average over time. It is worth noting that the measurement of second-order temporal correlation is only meaningful when the detection time response is less than or comparable to the coherence time of the light source.

(ii) Lévy statistics are described by the family of Lévy α-stable distributions, capturing the heavy-tailed nature of the intensity distribution and the emergence of extreme events, which is defined through the Fourier transform from the k-space of the characteristic function:

$$\overline{P}(k) = exp\{-|ck|^\alpha[1 - i\beta\ sgn(k)\Phi] + ikv\} \tag{2}$$

in which the Lévy index $\alpha \in (0, 2]$ is the most important parameter since, it governs the strength of the fluctuations.

(iii) Replica overlap parameter q and its distribution P(q), providing a macroscopic description phase transitions and revealing replica symmetry breaking (RSB) within the spin-glass framework, which can be expressed as:

$$q_{\alpha\beta} = \frac{\sum_{k=1}^{N} \Delta_\alpha(k)\Delta_\beta(k)}{\sqrt{\sum_{k=1}^{N} \Delta_\alpha^2(k)}\sqrt{\sum_{k=1}^{N} \Delta_\beta^2(k)}} \tag{3}$$

where the indexes $\alpha$ and k denote, respectively, the spectrum $\alpha$ emitted under identical experimental conditions (i.e., the replica $\alpha$) and the wavelength label; $\Delta_\alpha(k) = I_\alpha(k) - \overline{I}(k)$ and $\overline{I}(k)$ is the average over replicas of each wavelength intensity. Below we identify the parameter q as the locus (in absolute value) at which the distribution P(q) has its maximum.

These macroscopic descriptors collectively enable a unified physical picture linking photon statistics and phase transitions in RFLs. In contrast, conventional fiber lasers and ASE sources cannot support the full scope of this theoretical framework. Based on this foundation, the subsequent analysis will systematically investigate the photon-statistical evolution among the three types of sources (RFLs, conventional fiber lasers, and ASE) and establish a unified landscape of photon statistics and phase transitions in RFLs.

# Photon statistics

To uncover the distinctive photon-statistical behavior arising from disordered feedback, through theoretical modeling and experimental verification, this work systematically investigates three representative fiber-based light sources featuring fundamentally different feedback architectures: fixed-cavity single-frequency fiber lasers, feedback-free ASE sources, and Rayleigh-scattering-based RFL. These results enable us, for the first time, to construct a comprehensive "feedback mechanism–photon statistics" framework that unifies the photonic behaviors of resonant, feedback-free, and disordered light sources. Details of experimental setup and theoretical model are given in "Methods" and "Supplementary information I".

**Fixed-mirror-feedback and feedback-free light source.** To complement the study and for the sake of investigation, a commercial single-frequency fiber laser was selected as the single-frequency source. Its spectrum, shown in the Fig. 2a, exhibits a linewidth of less than 0.1 kHz. During the experiment, seven different pump powers were applied, and the second-order temporal correlation of the laser output was measured at each power level. As shown in Fig. 2b, the results demonstrate that the $g^{(2)}(0)$ remains constant at 1 across all pump powers. This indicates that the single-frequency fiber laser, based on a conventional resonant cavity, maintains stable second-order temporal correlations under varying pump conditions, consistently exhibiting highly correlated photon statistics expected for a coherent optical source.

As described in "Methods", we built a 1 $\mu m$ ASE source based on $Yb^{3+}$-ion gain and a 1.5 $\mu m$ ASE source based on $Er^{3+}$-ion gain. Under low pump power, the source operates primarily in the spontaneous emission regime. Due to their broad emission spectrum, the coherence time is much shorter than the response time of the detection system. Moreover, after applying spectral filtering, the signal of the spontaneous emission source will be drowned out by detector noise. Therefore, under such pump conditions, we investigated the second-order temporal correlation properties solely through simulation.

The statistical characteristics of the temporal intensity were analyzed and the corresponding second-order temporal correlation function was calculated. Fig. 2c and d show the simulated results of $g^{(2)}(\tau)$ and probability density function (PDF) of the spontaneous emission source under low pump power. The results indicate that the PDF of the temporal intensity follows an exponential distribution (consistent with the physical nature of ASE), and the second-order temporal correlation function $g^{(2)}(0) = 2$, demonstrating that the spontaneous emission source exhibits typical uncorrelated photon statistics.

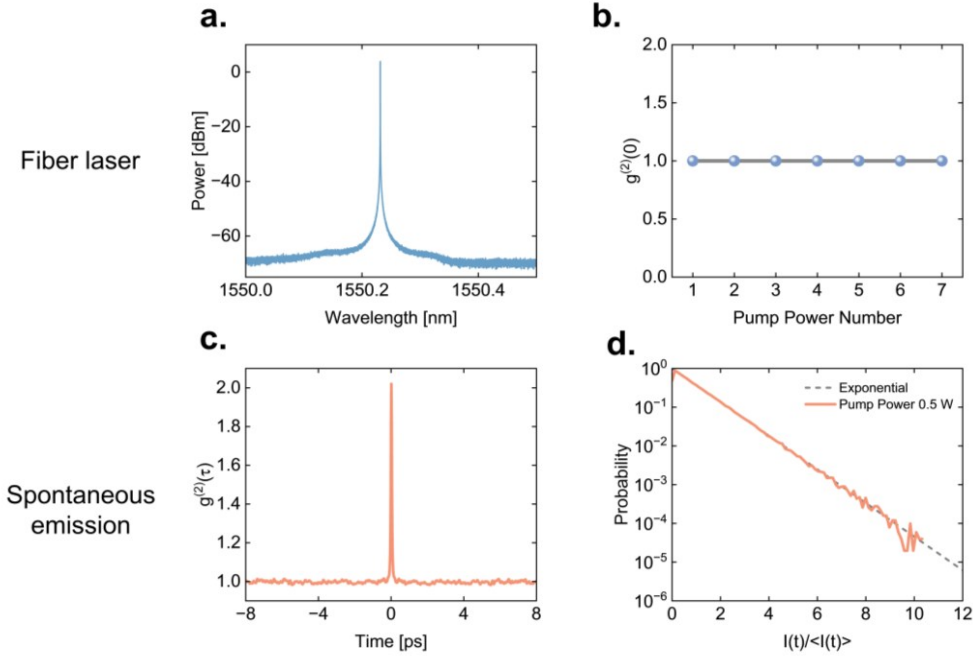

**Figure 2. Result of the single-frequency fiber laser and spontaneous emission source.** a. Spectrum of the single-frequency fiber laser; b. $g^{(2)}(0)$ of the single-frequency fiber laser under different pump powers: $g^{(2)}(0)$ remains constant at 1, which means that single-frequency fiber laser exhibits stable temporal correlations. c. $g^{(2)}(\tau)$ of the spontaneous emission source under pump power of 0.5 W: $g^{(2)}(0) = 2$ demonstrating that the spontaneous emission source exhibits typical uncorrelated photon statistics; d. PDF of the spontaneous emission source: following the exponential distribution under pump power of 0.5 W.

As the pump power increases, the spontaneous emission source gradually evolves into an ASE source. The evolution of the second-order temporal correlation of ASE source was investigated systematically through both experimental and simulation aspects. In the experiment, the temporal response of detection system is shorter than the coherence time of the source, satisfying the requirements for second-order temporal correlation measurements.

Figure 3 presents the evolution of the intensity PDF and $g^{(2)}(0)$ of the ASE sources under different pump powers, both in simulation and experiment. The results show that as the pump power increases, the intensity PDF gradually deviates from an exponential distribution, which means the intensity output becomes more stable. Simultaneously, the $g^{(2)}(0)$ value decreases, indicating enhanced coherence of the light field. The experimental results are consistent with the simulation.

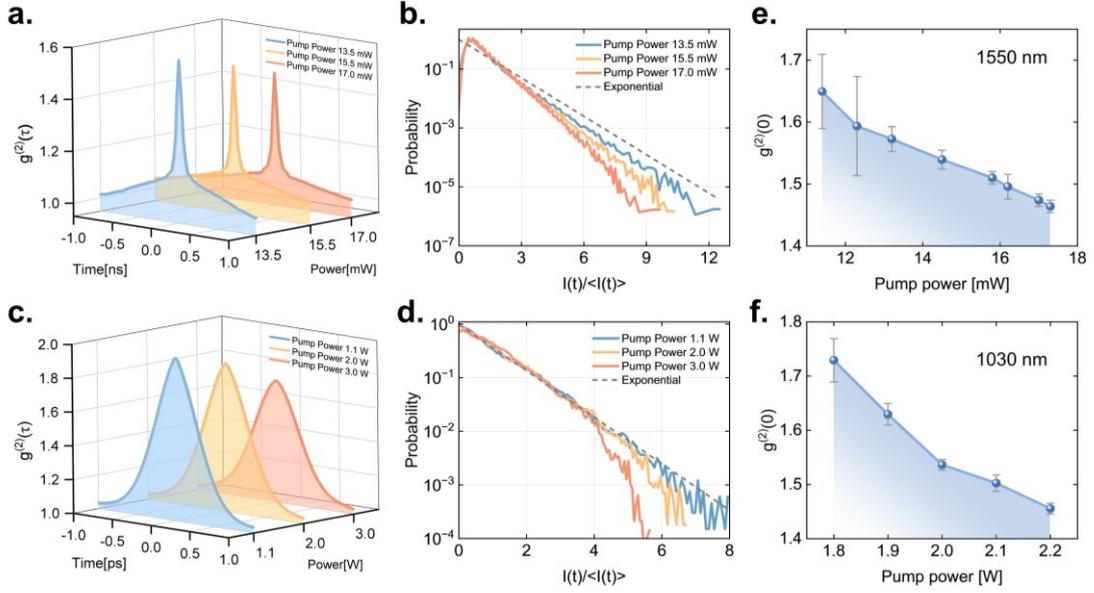

**Figure 3. Experimental and simulation results of the ASE source under different pump powers and wavelengths.** Under different pump power: Experimental results: a. $g^{(2)}(\tau)$ b. PDF of intensities. Simulation results: c. $g^{(2)}(\tau)$ d. PDF of intensities. With increasing pump power, $g^{(2)}(\tau)$ of the ASE gradually decreases, indicating an enhancement in temporal correlation. Correspondingly, the PDF progressively deviates inward from exponential distribution, suggesting an increasingly stable temporal intensity distribution. Experimental result under different wavelengths: e. 1550 nm; f. 1030 nm. ASE generated from different gain media and at different wavelengths exhibits the same evolutionary trend, further confirming the universality of this behavior in the ASE regime.

Furthermore, as shown in Fig. 3e and f, ASE sources constructed at different wavelengths in the experiment exhibit similar pump-power-dependent evolution trends in their second-order temporal correlation characteristics, demonstrating the universality of this phenomenon. The underlying physical mechanism is that, at low pump power, the emission is dominated by inherently incoherent spontaneous radiation. As the pump power increases, due to the non-uniformity of the gain spectrum, photons near the gain peak obtain more amplification. Consequently, the pump energy is preferentially converted into light at that spectral region, leading to spectral narrowing and the emergence of partial coherence. This finding enhances the understanding of the statistical properties and coherence evolution mechanisms of ASE sources.

**Randomly distributed-feedback light source.** In this work, a narrow-band Raman RFL based on Rayleigh scattering was constructed for the investigation of the evolution of photon statistics from both pump and scattering states. Regarding the different scattering states, firstly two experimental conditions were designed to study the influence of external perturbations on the photon statistics. In the first case (Case 1), the SMF was additionally protected to effectively isolate it from environmental disturbances. In the second case (Case 2), external stress was applied to the SMF to introduce perturbations. Previous studies have shown that Case 1 corresponds to relatively stable Rayleigh scattering phase variations within the fiber, whereas Case 2 is associated with significantly enhanced phase variations[27]. As shown in Fig. 4a, the output spectra under these two conditions exhibit distinct features[27]: in Case 1, the random laser spectrum contains multiple randomly distributed spikes, while in Case 2 these spikes disappear, and the spectrum exhibits a smoother envelope.

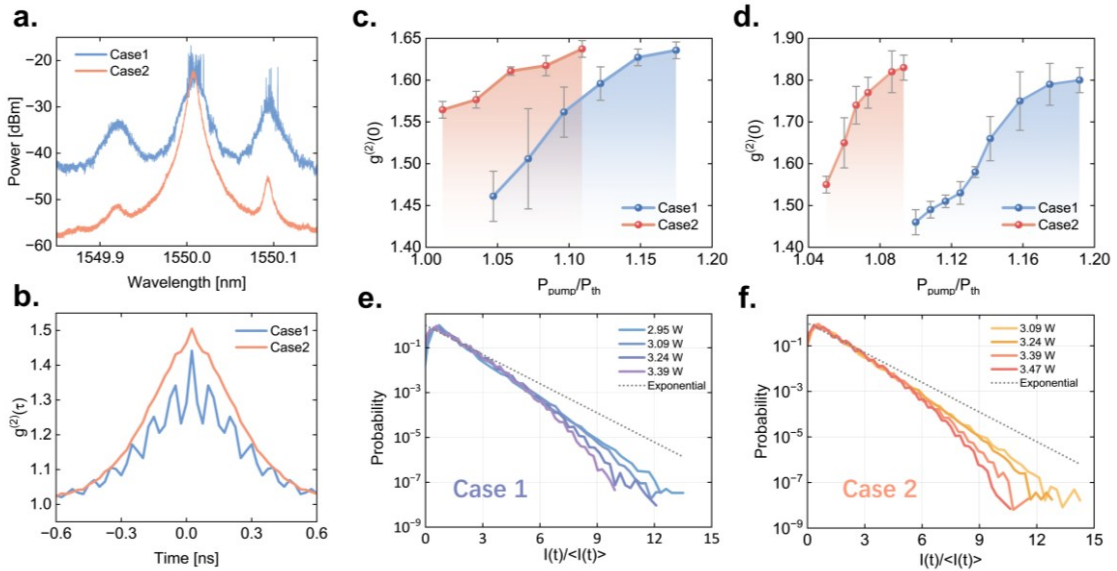

**Figure 4. Spectra, $g^{(2)}(\tau)$ and PDF of the RFL.** a. The spectra of the RFL in different cases. Case 1 corresponds to a stable Rayleigh scattering phase, whereas Case 2 features a fluctuating Rayleigh scattering phase. b. $g^{(2)}(\tau)$ of the RFL in the respective Cases 1 and 2. The periodic peaks observed in Case 1 correspond to the beat frequency between the main peak of the RFL and the stimulated Brillouin scattering (SBS) component. c. Experimental result of $g^{(2)}(0)$; d. Simulation result of $g^{(2)}(0)$. With increasing pump power, the $g^{(2)}(0)$ value of the RFL gradually increases, indicating a reduction in the second-order temporal correlation. Moreover, compared to the stable Rayleigh scattering phase, the fluctuating phase leads to higher $g^{(2)}(0)$ values and a weaker degree of second-order temporal correlation in the RFL output. PDF of the RFL: e. Case 1; f. Case 2. In contrast, with increasing pump power, the PDF of the RFL progressively deviates inward from the exponential

distribution, indicating a more stable temporal output. Notably, this evolutionary trend remains consistent across different Rayleigh scattering phase states, highlighting its universality in the RFL regime.

As shown in Fig. 4b, we calculated $g^{(2)}(\tau)$ of the laser output under two different experimental conditions. In Case 1, the spectrum of the RFL contains a stimulated Brillouin scattering (SBS) peak that is close in power to the main peak. The beat frequency interference between these two components leads to pronounced periodic oscillations in the $g^{(2)}(\tau)$ function, with a period corresponding to their frequency difference, which is 10 GHz (Case1). In contrast, in Case 2, the power difference between the SBS peak and the main peak is significantly small, resulting in negligible beat interference and the absence of periodic oscillations in $g^{(2)}(\tau)$. Furthermore, in Case 1, the presence of randomly distributed spectral spikes also contributes to additional beat interference among these spikes, giving rise to further periodic tenability in the $g^{(2)}(\tau)$.

The relationship between the second-order temporal correlation of the RFL and the pump power was investigated. Below the threshold, the RFL remains in the spontaneous emission regime with $g^{(2)}(0) = 2$, indicative of an uncorrelated state. Therefore, this investigation concentrates on the evolution of its second-order temporal correlations above threshold, experimentally and through simulations. As shown in Fig. 4c, d, e and f, both experimental and simulation results illustrate the evolution of the temporal intensity PDF and $g^{(2)}(0)$ as the pump power increases. Specifically, when the pump power exceeds the threshold, the RFL enters the random lasing regime. In this state, the temporal intensity PDF deviates from the exponential distribution, reflecting enhanced temporal-output stability. As the pump power continues to increase, this deviation becomes more pronounced, indicating increasingly stable temporal output. Correspondingly, when the pump power just reaches the threshold, $g^{(2)}(0)$ lies between 1 and 2, suggesting the emergence of coherence, and then gradually increases as the pump power rises further, indicating a reduction in coherence. This evolution trend suggests that the system exhibits strong coherence slightly above threshold, which gradually degrades at higher pump powers. Notably, this behavior contrasts sharply with that of ASE source under similar conditions, where $g^{(2)}(0)$ decreases monotonically with increasing pump power, and the intensity PDF consistently deviates from an exponential distribution, indicating improved output stability. Moreover, under both Case 1 and Case 2 experimental conditions, the RFL displays the same evolution pattern of $g^{(2)}(0)$.

The underlying physical mechanism of this phenomenon can be summarized as follows: When the pump power is below the threshold, the RFL operates in a spontaneous emission regime, exhibiting typical thermal light characteristics with strong temporal intensity fluctuations. In this state, the $g^{(2)}(0)$ = 2. As the pump power slightly above the threshold, the system enters the stimulated emission regime. However, the available pump energy is still limited and insufficient to simultaneously support multiple lasing modes, resulting in pronounced gain competition among them. This competition leads to significant intensity fluctuations in the time domain, with the presence of multiple extrema, and a strong photon correlation, causing a sharp drop in the $g^{(2)}(0)$ value. With further increases in pump power, the total available energy in the system rises, weakening the inter-mode gain competition. Consequently, the output becomes more stable, photon correlation weakens, and the temporal intensity PDF progressively deviates from the exponential distribution. Correspondingly, the $g^{(2)}(0)$ value gradually increases.

To gain deeper insight, the perturbation amplitude was subdivided to impose stresses of different strengths on the fiber, and the phase variations of Rayleigh scatterers were precisely measured using a phase-sensitive OTDR system with proprietary technologies (see Supplementary information II for details)[43]. As shown in Fig. 5a and b, increasing perturbation stress resulted in stronger phase variations, accompanied by a gradual suppression of randomly distributed spectral spikes, while the second-order temporal correlation exhibited a corresponding increase. These findings demonstrate that the second-order temporal correlations of RFLs undergo smooth evolution with enhanced scattering phase variations, thereby extending the controllable dimensions of their output characteristics.

# Unified Landscape for Photon Statistics and Photonic Phase Transitions

Previous studies have systematically investigated the intensity statistics in RFL and their associated photonic glassy phase transition phenomena, characterized by the Lévy index $\alpha$[44-46] and overlap parameter q (analogous to Parisi overlap parameter in spin-glass transition)[27,29,46], respectively. However, a unified exploration of photon statistics through the second-order temporal correlation function $g^{(2)}(0)$ and glassy phase transitions on a single optical platform has remained elusive, and the intrinsic links among these three aspects have yet to be established.

In this study, a Rayleigh-scattering-based RFL was employed as the experimental platform to systematically explore the intrinsic connections among these three aspects. Detailed investigations of the intensity statistics and the photonic phase transition are provided in Supplementary Materials III and IV.

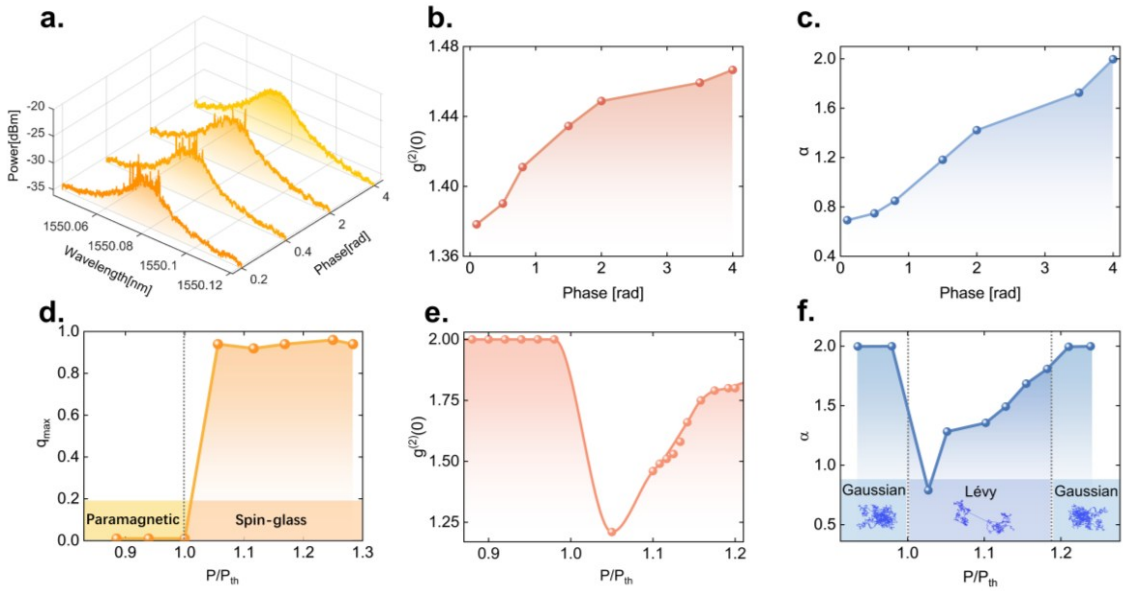

**Figure 5. q, $g^{(2)}(\tau)$ and $\alpha$ of RFL under different scattering phase states and pump power.** Scattering phase states: a. Spectra of the RFL; b. $g^{(2)}(0)$; **c.** Lévy index $\alpha$. The phase transition, intensity statistics, and second-order temporal correlations of RFL exhibit a unified evolution with Rayleigh scattering phase variations: as the fluctuation amplitude increases, the number of modes in the photonic spin-glass phase gradually decreases, the intensity statistics evolve toward a Gaussian distribution, and the second-order temporal correlation progressively weakens. Pump power: d. $q$; e. $g^{(2)}(0)$; f. Lévy index $\alpha$. All three cases exhibit a consistent evolutionary trend: Below the pump threshold, the RFL resides in a photonic paramagnetic phase ($q = 0$), with a $g^{(2)}(0)$ value equals 2 and a Gaussian intensity distribution ($\alpha = 2$). Near the threshold, the RFL transitions into a photonic RSB spin-glass phase ($q = 1$), where the $g^{(2)}(0)$ value drops toward 1 and the intensity distribution shifts to a Lévy regime ($0 < \alpha < 2$). With further increase in pump power, the $g^{(2)}(0)$ value rises again, and the intensity distribution gradually returns to a Gaussian distribution ($\alpha = 2$).

As summarized in Fig. 5d, e and f, the three observables ($\alpha$, q and $g^{(2)}(0)$) exhibit remarkably similar pump-power-dependent evolutions. Specifically, below threshold, $g^{(2)}(0)$ = 2, indicating temporally uncorrelated output. In parallel, the Lévy index $\alpha$ = 2, reflecting Gaussian statistics of independent modes, while the overlap parameter q = 0 corresponds to a photonic paramagnetic-like phase. Strikingly, just above threshold, all three parameters undergo abrupt changes: $g^{(2)}(0)$ drops below 2, revealing the onset of temporal correlations; the Lévy index $\alpha$ decreases below 2, signaling a transition to Lévy statistics in the frequency domain; and q = 1 denotes the emergence of a photonic RSB spin-glass phase. With further pump increase, both $g^{(2)}(0)$ and the Lévy index $\alpha$ gradually recover towards 2, consistent with the degradation of temporal correlations and the return to Gaussian intensity statistics.

Beyond the pump-power dimension, we also examined the role of scattering phase state, which had not been systematically explored before. By applying controlled perturbations to the fiber, the scattering phase was tuned into different fluctuation regimes, which were precisely measured using a phase-sensitive OTDR with proprietary technologies[43]. Previous studies suggested that above threshold, spectral spikes are in a photonic spin-glass phase, while smooth spectral envelopes are in a paramagnetic-like phase[27]. As shown in Fig. 5a, b and c, the three observables again display correlated trends with increasing scattering-phase variations: the number of spectral spikes diminishes, indicating a reduction in spin-glass modes, while both $g^{(2)}(0)$ and the Lévy index $\alpha$ increase, consistent with a progressive loss of temporal and spectral correlations.

The above results achieve, for the first time, a unified landscape for photon statistics and photonic phase transitions within the same experimental platform (Rayleigh-scattering-based RFL) across the dual dimensions of pump power and scatterer states. The findings demonstrate that these phenomena are not independent but are jointly driven by a common underlying mechanism. In contrast, such unification has never emerged in conventional light sources (e.g., single-frequency fiber lasers or ASE source), where the physical properties typically evolve in a mutually disconnected manner. For instance, single-frequency fiber lasers possess a single-valley energy landscape and reach a stable steady state once the pump exceeds threshold, leading to constant values of $g^{(2)}(0)$. ASE sources, characterized by a flat energy landscape, may show pump-dependent changes in second-order temporal correlations but do not undergo phase transitions or alterations of their underlying intensity statistics.

RFLs, however, benefit from the interplay among disorder, nonlinearity and amplification, enabling them to transcend the limitations of ordered photonic systems. As a result, photon statistics and phase transitions become integrated within a single platform, giving rise to highly cooperative macroscopic features. This work therefore fills a long-standing experimental gap in validating the unified behavior of disordered photonic systems and establishes RFLs as a paradigm platform for exploring statistical physics and critical phenomena in many-body complex systems.

## Discussion and Conclusion

This work systematically investigates the photon statistics of random fiber laser by second-order temporal correlation from both theoretical and experimental perspectives. First, by investigating the evolution of the $g^{(2)}(0)$ of three representative light sources (conventional single-frequency fiber laser, ASE, and RFL), the intrinsic differences in photon statistical properties under distinct feedback mechanisms are revealed. Specifically, single-frequency fiber laser exhibits highly coherent photon statistics that are independent of pump power, with a constant $g^{(2)}(0)$ value of 1. In contrast, ASE and RFL demonstrate pump-dependent dynamic evolution in both $g^{(2)}(0)$ and the PDF of temporal intensity. However, their trends are fundamentally opposite: with increasing pump power, the PDF of ASE gradually deviates from the exponential distribution, accompanied by a decrease in $g^{(2)}(0)$ from 2; conversely, in RFL, the PDF also deviates from the exponential form with increasing pump power, but the $g^{(2)}(0)$ value increases, signifying a degradation of temporal correlation. Furthermore, the study reveals that the second-order temporal correlation in RFL is influenced not only by pump power but also by the phase of Rayleigh scattering. Based on this insight, a two-dimensional tunability mechanism is proposed, providing a practical strategy for fine-tuning the output properties of RFL.

As a conclusion, this study, for the first time, achieves a unified landscape for photon statistics and photonic phase transitions within the same system. In the Rayleigh-scattering-based RFL, we find that these three aspects exhibit consistent evolutionary trends and transition boundaries along the dimensions of pump power and scattering states. Such unification, absent in conventional light sources, highlights the unique roles of disordered feedback and nonlinear effects in RFL, thereby establishing it as a paradigmatic model for exploring the statistical physics of complex systems and critical phenomena.

## Methods

**Experimental setup.** This study constructed an ASE source and narrow-linewidth RFL. Considering that commercial single-frequency fiber lasers are technologically mature and their theoretical models have been extensively validated, the reconstruction of such a system and further modeling were not undertaken. Instead, the commercial single-frequency fiber laser was employed as a benchmark reference, while the focus of this study lies in systematically acquiring and investigating the second-order temporal correlation functions across three distinct feedback regimes—single-frequency fiber laser, ASE, and RFL—to elucidate the fundamental differences in photon statistical behavior.

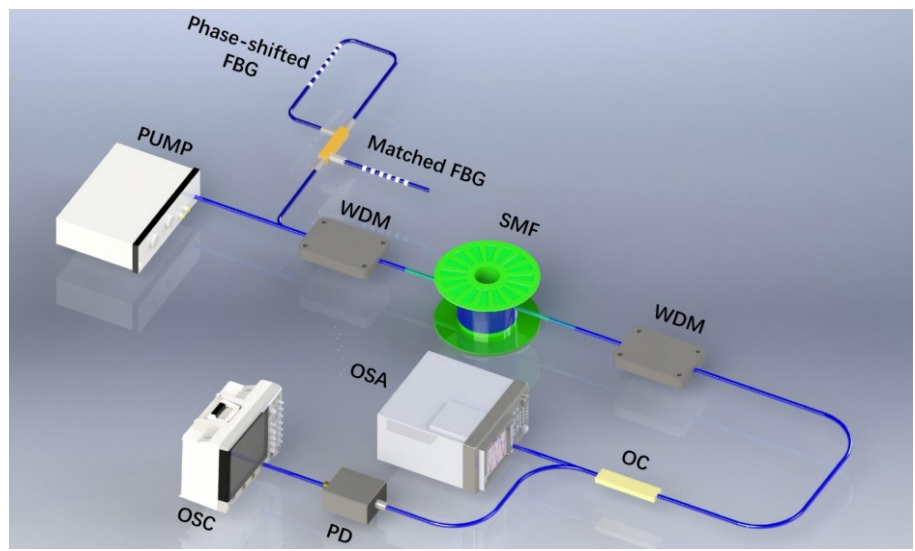

**Figure 7. Experiment setup of RFL.** FBG Fiber Bragg Grating; WDM wavelength division multiplexer; SMF single model fiber; OC optical coupler; PD photodetector; OSC oscilloscope; and OSA optical spectrum analyzer.

The experimental setup of the RFL is shown in Fig. 7. A 1455 nm pump was injected into a 15 km SMF via a 1455/1550 nm wavelength-division multiplexer (WDM). The 1550 nm port was connected to a feedback structure, which was provided by the transmission peak of a phase-shifted FBG and the reflection peak of a matched FBG. The generated RFL exhibited a spectral linewidth of 0.02 nm, corresponding to a coherence time of approximately $4 \times 10^{-10}$ s. Time-domain intensity fluctuations were recorded using a 40 GHz photodetector and a 16 GHz bandwidth oscilloscope, ensuring that the detection system's temporal resolution exceeded the coherence time of the laser source.

**Theoretical model for simulation results.** In this study, the theoretical modeling of ASE was focused on the 1 $\mu$m wavelength based on $Yb^{3+}$-ion gain, while the theoretical investigation of the RFL was conducted using a Rayleigh-scattering phase variation model[26].

### *ASE modeling*

The simulation model of ASE can be described as follows:

$$\pm\frac{\partial A^{\pm}}{\partial z}+\frac{i\beta_2}{2}\frac{\partial^2 A^{\pm}}{\partial t^2}+\frac{\alpha}{2}A^{\pm}=i\gamma\left(|A^{\pm}|^2+2|A^{\mp}|^2\right)A^{+}+g(|A^{\pm}|^2+|A^{\mp}|^2)A^{\pm} \quad (4)$$

$$g(P_s^+,P_s^-)=\frac{\alpha_p\mu\frac{P_p(z)}{P_{sat}^{(p)}}+\alpha_s\frac{P^+(z)+P^-(z)}{P_{sat}^{(s)}}}{1+\frac{P^+(z)+P^-(z)}{P_{sat}^{(s)}}+\frac{P_p(z)}{P_{sat}^{(s)}}} \quad (5)$$

$$\alpha_s=\Gamma_s\sigma_{as}N, \alpha_p=\Gamma_p\sigma_{ap}N \quad (6)$$

$$P_{sat}^{(s)}=\frac{h\nu_s A_c}{\tau(\sigma_{as}+\sigma_{es})\Gamma_s}, P_{sat}^{(p)}=\frac{h\nu_p A_c}{\tau(\sigma_{ap}+\sigma_{ep})\Gamma_p}, \mu^{-1}=\frac{\sigma_{es}}{\sigma_{ep}}\frac{\sigma_{ap}+\sigma_{ep}}{\sigma_{as}+\sigma_{es}} \quad (7)$$

where A is the complex envelope of the light-wave and P is the light-wave power, and in each iteration, random Gaussian white noise is set as the initial state to represent spontaneous emission light; $\alpha, \gamma, \beta_2$ are the linear fiber loss, Kerr coefficient and dispersion, respectively. The g is the gain of the YDF; $\Gamma_s$ and $\Gamma_p$ are the pump and emission overlap within the core; $\sigma_a$ and $\sigma_e$ are the absorption and emission cross-sections for the YDF. Based on this simulation model, the output characteristics of ASE is obtained.

### *RFL modeling*

The Rayleigh-scattering phase variation model of RFL can be expressed as:

$$\frac{\partial u_p^{\pm}}{\partial z}\mp\frac{1}{v_{gs}}\frac{\partial u_p^{\pm}}{\partial t}\pm i\frac{\beta_{2p}}{2}\frac{\partial^2 u_p^{\pm}}{\partial t^2}\pm\frac{\alpha_p}{2}u_p^{\pm}=\pm i\gamma_p|u_p^{\pm}|^2u_p^{\pm}\mp\frac{g_p(\omega)}{2}\left(\langle|u_s^{\pm}|^2\rangle+\langle|u_s^{\mp}|^2\rangle\right)u_p^{\pm} \quad (8)$$

$$\frac{\partial u_s^{\pm}}{\partial z}\pm i\frac{\beta_{2s}}{2}\frac{\partial^2 u_s^{\pm}}{\partial t^2}\pm\frac{\alpha_s}{2}u_s^{\pm}\mp\frac{\varepsilon(\omega,t)}{2}u_s^{\pm}=\pm i\gamma_s|u_s^{\pm}|^2u_s^{\pm}\pm\frac{g_s(\omega)}{2}\left(\langle|u_p^{\pm}|^2\rangle+\langle|u_p^{\mp}|^2\rangle\right)u_s^{\pm} \quad (9)$$

where the subindexes 'p' and 's' represent the pump wave and stokes wave respectively; "+" and "-" correspond to forward and backward light; $u, v_{gs}, \omega$ are the envelop of the optical field, group velocity difference and the angular frequency of light-wave, respectively; $g$ is the Raman gain. In addition, different from the ASE model, the Rayleigh-scattering phase variation model incorporates a Rayleigh scattering term, as its output is critically determined by the feedback mechanism: ε is the Rayleigh scattering, which is a time-varying term, and its time-varying intervals of frequency and phase depend on the specific experimental environment. In this simulation, the value of the Rayleigh scattering term is measured precisely using a Φ-OTDR system with proprietary technologies.

## Acknowledgements

This work was supported by the National Natural Science Foundation of China 62435002 (ZNW), 62075030 (ZNW), Ministry of Science and Technology of China DL2023167001L (ZNW), Sichuan Science and Technology Program under 2023YFSY0058 (ZNW), 111 Project B14039 (ZNW). A.S.L.G. and E.P.R. thank Brazilian funding agencies CNPq, FACEPE and INFo.

**Author Contributions.** Z. W. conceived the idea of this study. Y. Q., J. L., and T. W. developed the numerical simulations. Y. Q., R. L. and W. L. carried out the experiment. Y. Q., E. P. R., A. S. L. G, and Z. W. wrote the manuscript and analyzed the results. All the authors discussed the results and contributed to the manuscript.

**Conflict of interest.** Authors declare that they have no competing interests.

**Data and materials availability.** All data are available in the main text or the supplementary materials.

**Supplementary Information.** All copyrights are reserved in www.nature.com.

## References：

1. Parisi, G. Nobel Lecture: Multiple equilibria. *Reviews of modern physics* **95**, 030501 (2023).

2. Parisi, G. Infinite number of order parameters for spin-glasses. *Physical Review Letters* **43**, 1754–1756 (1979).

3. The power of fluctuations. *Nature Physics* **17**, 1185 (2021).

4. Buchhold, M. et al. Dicke-model quantum spin and photon glass in optical cavities:

nonequilibrium theory and experimental signatures. *Physical Review A* **87**, 063622 (2013).

5. Sauerwein, N. et al. Engineering random spin models with atoms in a high-finesse cavity. *Nature Physics* **19**, 1128–1134 (2023).

6. Turitsyn, S. K. et al. Random distributed feedback fibre laser. *Nature Photonics* **4**, 231-235 (2010).

7. Wiersma, D. The physics and applications of random lasers. *Nature Physics* **4**, 359–367 (2008).

8. Churkin, D. V. et al. Raman fiber lasers with a random distributed feedback based on Rayleigh scattering, *Physical Review A* **82**, 033828 (2010).

9. Qi, Y. F. et al. Statistical physics and nonlinear dynamics in random fiber lasers: from theory to multidisciplinary applications[J]. *Advanced Photonics*, **8(1)**, 014001 (2026).

10. Gomes, A. S. L. et al. Recent advances and application of random lasers and random fiber lasers, *Progress in Quantum Electronics* **78**, 100343 (2021).

11. Churkin, D. V. et al. Wave kinetics of random fibre lasers, *Nature Communications* **6**, 6214 (2015).

12. Sugavanam, S. et al. Spectral correlations in a random distributed feedback fibre laser, *Nature Communications* **8**, 15514 (2017).

13. Mercadier, N. et al. Lévy flights of photons in hot atomic vapours, *Nature Physics* **5**, 602-605 (2009).

14. Alves, N. P. et al. Observation of Replica Symmetry Breaking in Standard Mode-Locked Fiber Laser. *Physics Review Letter* **132**, 093801 (2024).

15. González, I. R. R. et al. Coexistence of turbulence-like and glassy behaviours in a photonic system. *Scientific Reports* **8**, 17046 (2018).

16. Gomes, A.S.L. et al. Lévy Statistics and Spin Glass Behavior in Random Lasers 1st ed. (Jenny Stanford Publishing, 2023).

17. Wang, Z. N. et al. High power random fiber laser with short cavity length: theoretical and experimental investigations. *IEEE Journal of Selected Topics in Quantum Electronics* **21**, 10-15 (2015).

18. Zhang, Y. et al. Spectrally programmable Raman fiber laser with adaptive wavefront

shaping, *Photonics Research* **11**, 20-26 (2023).

19. Li, S. et al. Multi-wavelength random fiber laser with a spectral-flexible characteristic, *Photonics Research* **11**, 159-164 (2023).

20. Tan, M. et al. Transmission performance improvement using random DFB laser based Raman amplification and bidirectional second-order pumping, *Optics Express* **24**, 2215-2221 (2016).

21. Jia, X. H. et al. Random-lasing-based distributed fiber-optic amplification, *Optics Express* **21**, 6572-6577 (2013).

22. Redding, B. et al. Speckle-free laser imaging using random laser illumination, *Nature Photonics* **6**, 355-359 (2012).

23. Ma, R. et al. Multimode random fiber laser for speckle-free imaging, *IEEE Journal of Selected Topics in Quantum Electronics* **25**, 1-6 (2019).

24. Lin, S. T. et al. Wideband remote-sensing based on random fiber laser, *Journal of Lightwave Technology* **40**, 3104-3110 (2022).

25. Qi, Y. F. et al. Impact of feedback bandwidth on Raman random fiber laser remote-sensing, *Optics Express* **30**, 21268-21275 (2022).

26. Fan, M. Q. et al. Spectrum-tailored random fiber laser towards ICF laser facility. *Matter and Radiation at Extremes* **8**, 025902(2023).

27. Qi, Y. F. et al. Replica symmetry breaking in 1D Rayleigh scattering system: theory and validations, *Light: Science & Applications* **13**, 151 (2024).

28. Du, W. et al. Observation of the photonic hall effect and photonic magnetoresistance in random lasers, *Nature Communications* **15**, 4589 (2024).

29. Ghofraniha, N. et al. Experimental evidence of replica symmetry breaking in random lasers, *Nature Communications* **6**, 6058 (2015).

30. González, I. R. R. et al. Turbulence hierarchy in a random fibre laser, *Nature Communications* **8**, 15731 (2017).

31. Conti, C. & DelRe, E. Photonics and the Nobel Prize in physics, *Nature Photonics* **16**, 6-7 (2022).

32. Gomes, A. S. L. et al. Photonics bridges between turbulence and spin glass phenomena in the 2021 Nobel Prize in Physics. *Light*: *Science & Applications* **11**, 104

(2022).

33. Moura, A. L. et al. Nonlinear effects and photonic phase transitions in $Nd^{3+}$-doped nanocrystal-based random lasers, *Applied Optics* **59**, D155-D162 (2020).

34. Moura, A. L. et al. Replica symmetry breaking in the photonic ferromagneticlike spontaneous mode-locking phase of a multimode Nd:YAG laser, *Physical Review Letters* **119**, 163902 (2017).

35. Boitier, F. Et al. Measuring photon bunching at ultrashort timescale by two-photon absorption in semiconductors, *Nature Physics* **5**, 267-270 (2009).

36. BROWN, R. H. & TWISS, R. Q. Correlation between photons in two coherent beams of light, *Nature* **177**, 27-29 (1956).

37. Loudon, R. The Quantum Theory of Light (Oxford: Oxford University Press, 2000).

38. Ihn, Y. S. et al. Second-order temporal interference with thermal light: Interference beyond the coherence time, *Physical Review Letters* **119**, 263603 (2017).

39. Cao, H. et al. Photon statistics of random lasers with resonant feedback, *Physical Review Letters* **86**, 4524 (2001).

40. Florescu, L. & John, S. Photon statistics and coherence in light emission from a random laser, *Physical Review Letters* **93**, 013602 (2004).

41. González, I. R. R. et al. Intensity distribution in random lasers: comparison between a stochastic differential model of interacting modes and random phase sum-based models, *Journal of the Optical Society of America B-Optical Physics* **38**, 2391-2398 (2021).

42. Raposo, E. P. et al. Intensity $g^{(2)}$ correlations in random fiber lasers: A random-matrix-theory approach, *Physical Review A* **105**, L031502 (2022).

43. Jiang, J. L. et al. Continuous chirped-wave phase-sensitive optical time domain reflectometry. *Optics Letters* **46**, 685-688 (2021).

44. Lima, B. C. et al. Observation of Lévy statistics in one-dimensional erbium-based random fiber laser, *Journal of the Optical Society of America B-Optical Physics* **34**, 293-299 (2017).

45. Li, J. et al. Lévy spectral intensity statistics in a Raman random fiber laser, *Optics Letters* **44**, 2799-2802 (2019).

46. Gomes, A. S. L. et al. Glassy behavior in a one-dimensional continuous-wave erbium-doped random fiber laser. *Physical Review A* **94**, 011801 (2016).